# Fast recursive filters for simulating nonlinear dynamic systems

**J. H. van Hateren**

j.h.van.hateren@rug.nl

*Netherlands Institute for Neuroscience, Royal Netherlands Academy of Arts and Sciences, Amsterdam, and Institute for Mathematics and Computing Science, University of Groningen, The Netherlands*

**Abstract**

A fast and accurate computational scheme for simulating nonlinear dynamic systems is presented. The scheme assumes that the system can be represented by a combination of components of only two different types: first-order low-pass filters and static nonlinearities. The parameters of these filters and nonlinearities may depend on system variables, and the topology of the system may be complex, including feedback. Several examples taken from neuroscience are given: phototransduction, photopigment bleaching, and spike generation according to the Hodgkin-Huxley equations. The scheme uses two slightly different forms of autoregressive filters, with an implicit delay of zero for feedforward control and an implicit delay of half a sample distance for feedback control. On a fairly complex model of the macaque retinal horizontal cell it computes, for a given level of accuracy, 1-2 orders of magnitude faster than 4th-order Runge-Kutta. The computational scheme has minimal memory requirements, and is also suited for computation on a stream processor, such as a GPU (Graphical Processing Unit).

**1 Introduction**

Nonlinear systems are ubiquitous in neuroscience, and simulations of concrete neural systems often involve large numbers of neurons or neural components. In particular if model performance needs to be compared with and fitted to measured neural responses, computing times can become quite restrictive. For such applications, efficient computational schemes are necessary. In this article, I will present such a highly efficient scheme, that has recently been used for simulating image processing by the primate outer retina (van Hateren 2006, 2007). The scheme is particularly suited for data-driven applications, where the time step of integration is dictated by the sampling interval of the analog-to-digital or digital-to-analog conversion. It assumes that the system can be decomposed into components of only two types: static nonlinearities and first-order low-pass filters. Interestingly, these components are also the most common ones used in neuromorphic VLSI circuits (Mead 1989). In the scheme presented here, the components need not have fixed parameters, but are allowed to depend on the system state. They are arranged in a possibly complex topography, typically involving several feedback loops. The efficiency of the scheme is produced by using very fast recursive filters for the first-order low-pass filters. I will show that it is best to use slightly different forms of the filter algorithm for feedforward and feedback processing loops.

No attempt is made to rigorously analyze convergence or optimality of the scheme, which would anyway be difficult to do for arbitrary nonlinear systems. The scheme should therefore be viewed as a practical solution, that works well for the examples I give in this article, but may need specific testing and benchmarking on new problems.

The scheme I present here can be efficiently implemented on stream processors. Recently there has been growing interest in using such processors for high performance computing (e.g., Göddeke et al. 2007, Ahrenberg et al. 2006, Guerrero-Rivera et al. 2006). In particular the arrival of affordable graphical processing units (GPUs) with raw computating power more than an order of magnitude higher than that of CPUs is driving this interest (see http://www.gpgpu.org). Current GPUs typically have about 100 processors that can work in parallel on data in the card's memory. Once the data and the (C-like) programs are loaded into the card, the card computes essentially independently of the CPU. Results can subsequently be uploaded to the CPU for further processing. GPUs are especially suited for simulating problems, such as in retinal image processing, that can be written as parallel, local operations on a two-dimensional grid.

Stream processors are, unlike CPUs, data driven and not instruction driven. They process the incoming data as it becomes available, and therefore usually need algorithms with fixed, or at least predictable computing times. The processing scheme I present in this article has indeed a fixed computing time. Moreover, it has low computational cost and low memory requirements, because it only deals with current and previous values of input, state variables, and output. The output is produced without delays that are not part of the model, that is, at the same time step as the current input, and the scheme is thus also suited for real-time applications.

The article is organized as follows. First, I will present a fairly complete overview of methods to simulate a first-order low-pass filter with a minimal recursive filter. Subsequently, I will give several examples of how specific neural systems - in particular several subsystems of retinal processing and spike generation following the Hodgkin-Huxley equations - can be decomposed into suitable components. Computed results of the various forms of recursive filters are compared with benchmark calculations using a standard Matlab solver. It is shown that for a practical, fairly complex model the most efficient algorithm (modified Tustin) outperforms a conventional 4th-order Runge-Kutta integration by 1-2 orders of magnitude. Finally, I will discuss the merits and limits of the approach taken here.

**2 Discrete simulation of a first-order low-pass filter**

Much of the material presented in this section is not new. However, I found that most of it is scattered throughout the literature, and I will therefore give a fairly complete overview. Table 1 summarizes the filters and their properties.

In the continuous time domain, the equation

$$\frac{dy}{dt} + \frac{1}{\tau} y = \frac{1}{\tau} x, \tag{1}$$

describes a first-order low-pass filter transforming an input function $x(t)$ into an output function $y(t)$, where $\tau$ is the time constant, and the coefficient in front of $x$ is chosen such that the filter has unit DC gain: $y=x$ if the input is a constant. In the examples below, I will usually write this equation in the standard form

$$\tau \dot{y} = x - y. \tag{2}$$

Fourier transforming this equation gives as the transfer function of this filter

$$H(\omega) = \frac{\tilde{y}}{\tilde{x}} = \frac{1}{1 + i\omega\tau}, \tag{3}$$

where the tilde denotes Fourier transforms. The impulse response of the filter is

$$\begin{aligned} h(t) &= \frac{1}{\tau} e^{-t/\tau} \quad \text{for } t \geq 0 \\ &= 0 \quad \quad \text{for } t < 0. \end{aligned} \tag{4}$$

We will assume now that $x(t)$ is only available at discrete times $t_n = n\Delta$, as $x_n = x(n\Delta)$, and that we only require $y(t)$ at the same times, as $y_n = y(n\Delta)$. Here $\Delta$ is the time between samples. Conforming with the most common integration schemes, we will further assume that for calculating the current value of the output only the current value of the input, the previous value of the output, and possibly the previous value of the input are available. We therefore seek real coefficients $a_1$, $b_0$, and $b_1$ such that

$$y_n = -a_1 y_{n-1} + b_0 x_n + b_1 x_{n-1} \tag{5}$$

produces an output close to that expected from Eq. (2). The indices and signs of the coefficients are chosen here in such a way that they are consistent with common use in the digital processing community for describing IIR (infinite impulse response) or ARMA (auto-regressive, moving average) filters that relate the z-transforms of input and output (Oppenheim and Schafer 1975). I will not use the z-transform formalism here, but only note that Fourier transforming Eq. (5) and using the shift theorem gives

$$\tilde{y}_n = -a_1 \tilde{y}_n e^{-i\omega\Delta} + b_0 \tilde{x}_n + b_1 \tilde{x}_n e^{-i\omega\Delta}, \tag{6}$$

and therefore a transfer function

$$H(\omega) = \frac{\tilde{y}_n}{\tilde{x}_n} = \frac{b_0 + b_1 z^{-1}}{1 + a_1 z^{-1}}, \tag{7}$$

where the operator $z^{-1} = \exp(-i\omega\Delta)$ represents a delay of one sample.

The coefficients $a_1$, $b_0$, and $b_1$ are not independent because of the additional constraint that the filter of Eq. (2) has unit DC gain. A constant input $c$ must then produce a constant output $c$, thus Eq. (5) yields $c = -a_1 c + b_0 c + b_1 c$ and therefore

$$-a_1 + b_0 + b_1 = 1. \tag{8}$$

Because representing a general continuous system as in Eq. (2) by a discrete system as in Eq. (5) can only be approximate (note that Eqs. 3 and 7 cannot be made identical), there is no unique choice for the coefficients $a_1$, $b_0$, and $b_1$. Below I will give an overview of several possibilities, mostly available in the literature, and discuss their appropriateness for the computational scheme to be presented below. The first three methods discussed below, forward Euler, backward Euler, and the Trapezoidal rule, are derived from general methods for approximating derivatives. The further methods discussed are more specialized, dealing specifically with Eq. (2) and differing with respect to how the input signal is assumed to behave between the sampled values.

**2.1 Forward Euler**

Forward Euler (Press et al. 1992) is quite often used in neural simulations. Applied to Eq. (2) it amounts to the approximation

$$y_n \approx y_{n-1} + \dot{y}_{n-1} \Delta = y_{n-1} + (x_{n-1} - y_{n-1})\Delta/\tau, \tag{9}$$

hence we get the recurrence equation

$$\begin{aligned} y_n &= (1 - 1/\tau') y_{n-1} + (1/\tau') x_{n-1} \\ &\text{with} \quad \tau' = \tau/\Delta. \end{aligned} \tag{10}$$

Here as well as below I will use $\tau'$, which is $\tau$ normalized by the sample distance, to keep the equations concise. Eq. (10) suffers from two major problems: first, it is not very accurate, and even unstable for small $\tau'$ (Press et al. 1992), and second, it produces an implicit delay of $\Delta/2$ for centered samples. The second problem is illustrated in Fig. 1. Figure 1A shows a starting sinusoid, where the filled circles give the function values at the sampling times. The continuous function of Fig. 1A can subsequently be filtered by Eq. (2) using a standard integration routine (Matlab ode45) at a time resolution much better than $\Delta$ (obviously, in this simple case the result could have been obtained analytically, but we will encounter other examples below where this is not possible). Fig. 1B shows the result (continuous line). When the samples of the sinusoid are processed by Eq. (10), the result lags by half a sampled distance (red open circles in Fig. 1B).

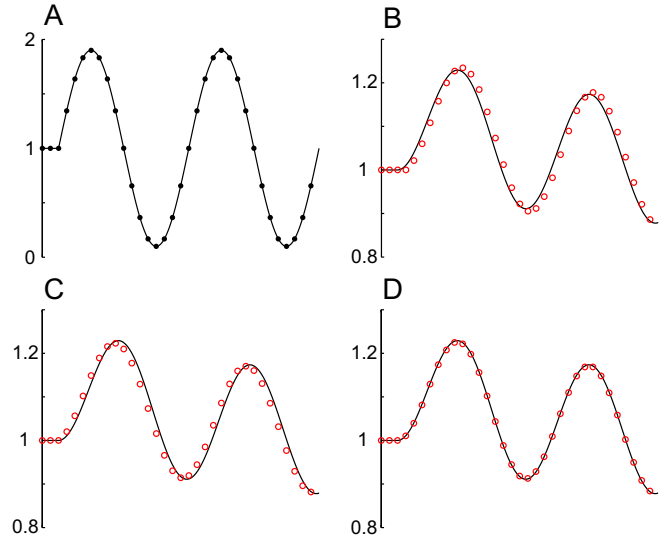

Figure 1. (A) Starting sinusoid (continuous line) and function values at the sample times (filled circles, 16 samples per period). The function equals 1 at times earlier than shown. (B) Continuous line: sinusoid of (A) filtered by Eq. (2) with $\tau'=16$, computed with Matlab ode45; red open circles: result of filtering the samples of (A) with Eq. (10), the recurrence equation that follows from forward Euler. Output samples lag by approximately half a sample distance. (C) As (B), for backward Euler (Eq. (12)). Output samples lead by approximately half a sample distance. (D) As (B), for Trapezoidal (Eq. (14)).

## 2.2 Backward Euler

Backward Euler (Press et al. 1992) applied to Eq. (1) yields
$$y_n \approx y_{n-1} + \dot{y}_n \Delta = y_{n-1} + (x_n - y_n)\Delta/\tau, \tag{11}$$
hence
$$y_n = [\tau'/(\tau'+1)]y_{n-1} + [1/(\tau'+1)]x_n. \tag{12}$$

Backward Euler is stable (Press et al. 1992) and slightly more accurate than forward Euler, but suffers from the problem that it produces an implicit delay of $-\Delta/2$ for centered samples, that is, a phase advance. Fig. 1C illustrates this, where the continous curve is the correct result (identical curve as the black curve in Fig. 1B), and the red open circles give the result of applying Eq. (12).

## 2.3 Trapezoidal rule

The trapezoidal rule (also known as Crank-Nicholson, Rotter and Diesmann 1999) is equivalent to the bilinear transformation and Tustin's method in digital signal processing (Oppenheim and Schafer 1975). It combines forward and backward Euler:
$$y_n \approx y_{n-1} + \tfrac{1}{2}(\dot{y}_{n-1} + \dot{y}_n)\Delta = y_{n-1} + \tfrac{1}{2}(x_{n-1} - y_{n-1} + x_n - y_n)\Delta/\tau, \tag{13}$$
and leads to
$$y_n = [(\tau'-0.5)/(\tau'+0.5)]y_{n-1} + [0.5/(\tau'+0.5)]x_n + [0.5/(\tau'+0.5)]x_{n-1}. \tag{14}$$

The method is stable, accurate, and produces a negligible implicit delay (Fig. 1D).

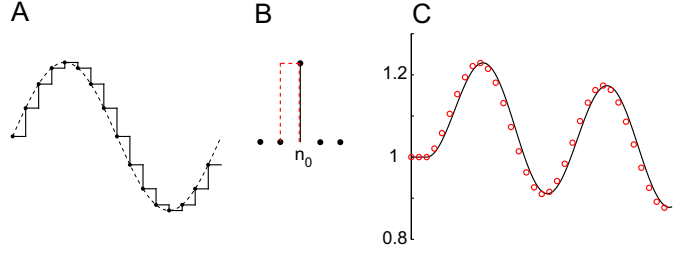

Figure 2. (A) Zero-Order Hold sampling model, where the sample values (dots) taken from a function (dashed line) are hold until a new sample arrives (continuous line). (B) A unit sample (black line and filled circle) is assumed here to represent a block in the previous inter-sample interval (red dashed line) (C) Continuous line: sinusoid of Fig. 1A filtered by Eq. (2) with τ'=16, computed with Matlab ode45; red open circles: result of filtering the samples of Fig. 1A with Eq. (17), the recurrence equation that follows from the ZOH processing scheme (i.e., assumed pulse shape of (B)).

## 2.4 Exponential Euler

A method that has gained some popularity in the field of computational neuroscience (for example in the simulation package Genesis, Bower and Beeman 1998) is sometimes called Exponential Integration (MacGregor 1987, Rotter and Diesmann 1999) or Exponential Euler (Moore and Ramon 1974, Rush and Larsen 1978, Butera and McCarthy 2004). It assumes that the input is approximately constant, namely equal to $x_{n-1}$, on the interval from $(n-1)\Delta$ to $n\Delta$. Equation (1) then has the exact solution (see e.g. appendix C.6 of Rotter and Diesmann 1999)

$$y_n = e^{-1/\tau'} y_{n-1} + (1 - e^{-1/\tau'}) x_{n-1}. \qquad (15)$$

This method is closely related to forward Euler, as a comparison of Eqs. (10) and (15) shows: for large $\tau'$ (time constant large compared with the sample distance), the factors $\exp(-1/\tau') \approx 1 - 1/\tau'$ and $1 - \exp(-1/\tau') \approx 1/\tau'$ approximate those of forward Euler. The exponential Euler method is stable, and more accurate than forward Euler for small $\tau'$. However, it has the same implicit delay of $\Delta/2$ as forward Euler (not shown).

## 2.5 Zero-Order Hold (ZOH)

When using analog-to-digital and digital-to-analog converters, a choice has to be made for the assumed signal values between the sample times. A simple practical choice is to keep the value of the last sample until a new sample arrives. This is called a zero-order hold (ZOH), and for a sampled sinusoid it assumes the continuous line shown in Fig. 2A. It involves an implicit delay of $\Delta/2$. Digitally filtering the samples of a ZOH system can compensate for this delay by assuming that a unit sample at $n = n_0$ (black line and filled circle in Fig. 2B) represents a block as shown by the dashed red line in Fig. 2B. The coefficients $a_1$, $b_0$, and $b_1$ for approximating Eq. (2) by Eq. (5) can be readily obtained from the response to this pulse; these coefficients then also apply to an arbitrary input signal, because the filter is linear and time-invariant. For samples $n \geq n_0 + 2$, the present and previous input are zero, thus the terms with $b_0$ and $b_1$ do not contribute. Because Eq. (4) shows that the output must decline exponentially, we find $-a_1 = e^{-\Delta/\tau} = e^{-1/\tau'}$. For sample $n = n_0$, the previous input and output are zero, thus the terms with $a_1$ and $b_1$ do not contribute. We then find $b_0$ from the convolution of the block $s(t)$ (dashed line in Fig. 2B) with the pulse response $h(t)$ of the filter, evaluated at sample $n = n_0$

$$b_0 = \left.\int_{-\infty}^{\infty} h(t')p(t-t')dt'\right|_{t=n_0\Delta} = \int_0^\Delta \frac{1}{\tau} e^{-t'/\tau} \cdot 1 dt' = 1 - e^{-\Delta/\tau} = 1 - e^{-1/\tau'}. \quad (16)$$

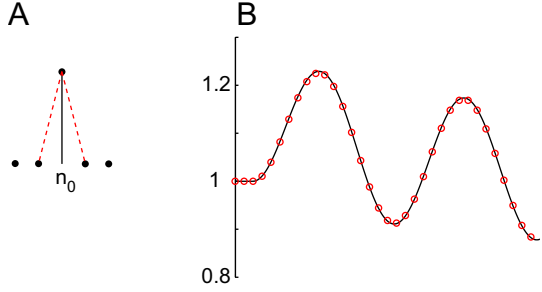

Figure 3. (A) A unit sample (black line and filled circle) is assumed here to represent linear interpolation in the previous and next inter-sample intervals (red dashed line) (B) Continuous line: sinusoid of Fig. 1A filtered by Eq. (2) with τ'=16, computed with Matlab ode45; red open circles: result of filtering the samples of Fig. 1A with Eq. (19), the recurrence equation that follows from the FOH processing scheme (i.e., assumed pulse shape of (A)).

With Eq. (8) we then find $b_1 = 1 + a_1 - b_0 = 0$. The recurrence equation therefore is

$$y_n = e^{-1/\tau'} y_{n-1} + (1 - e^{-1/\tau'}) x_n. \quad (17)$$

Note that the difference with Eq. (15) is that here the current input sample, $x_n$, is used, where in Eq. (15) it is the previous input sample, $x_{n-1}$. Whereas Eq. (15) implies a delay of $\Delta/2$, the present scheme has a delay of $-\Delta/2$, i.e., a phase advance (see Fig. 2C).

The filter in Eq. (17) is a special case of a general scheme of representing linear filters by using the matrix exponential (e.g., Rotter and Diesmann 1999, where it is called Exact Integration). Such filters are consistent with assuming a ZOH, and therefore imply a delay of $-\Delta/2$. Although Rotter and Diesmann (1999) do not use a ZOH but a function representation using Dirac δ-functions, a delay is implied by the choice of integration interval in their Eq. (3), which excludes the previous input sample and fully includes the present input sample. Had the integration interval been chosen symmetrical, the δ-functions at the previous and present input samples would each have contributed by one half, leading to a scheme with $0.5(x_{n-1} + x_n)$ as input, and therefore an implicit delay of 0.

**2.6 First-Order Hold (FOH)**

Another choice for the assumed function values between samples is the first-order hold (FOH), where sample values are connected by straight lines. It assumes that a unit sample at $n = n_0$ (black line and filled circle in Fig. 3A) represents a triangular pulse as shown by the dashed red line in Fig. 3A. The method is also called the triangular or ramp-invariant approximation, and is in fact equivalent to assuming that a function can be represented by B-splines of order one (Unser 1999, 2005). A general derivation of the recurrence relation, also valid for the more general lead-lag system $\tau_y \dot{y} + y = \tau_x \dot{x} + x$ of which Eq. (2) is a special case, is given by Brown (2000). A simple, alternative derivation goes similarly as given above for the ZOH. For samples $n \geq n_0 + 2$, the present and previous input are zero, and we again find $-a_1 = e^{-1/\tau'}$. For sample $n = n_0$, the previous input and output are zero, and now $b_0$ equals

$$b_0 = \int_{-\infty}^{\infty} h(t')p(t-t')dt' \bigg|_{t=n_0\Delta} = \int_0^{\Delta} \frac{1}{\tau} e^{-t'/\tau}(1-\frac{t'}{\Delta})dt' = 1 - \tau' + \tau' e^{-1/\tau'}. \qquad (18)$$

With Eq. (8) we then find $b_1 = 1 + a_1 - b_0 = \tau' - (1+\tau')\exp(-1/\tau')$. The recurrence equation therefore is

$$y_n = e^{-1/\tau'} y_{n-1} + (1 - \tau' + \tau' e^{-1/\tau'})x_n + (\tau' - (1+\tau')e^{-1/\tau'})x_{n-1}. \qquad (19)$$

Fig. 3B illustrates that the FOH has a negligible implicit delay.

**2.7 Centered Step-Invariant**

The centered step-invariant approximation (e.g., Thong and McNames 2002) is not often used, and is given here only for completeness; its performance is similar to that of FOH and Trapezoidal. It assumes that a unit sample at $n = n_0$ represents a block that is, contrary to the regular zero-order hold, centered on the sample time. This is equivalent to assuming that a function can be represented by B-splines of order zero (Unser 1999). As before, we must have $-a_1 = e^{-1/\tau'}$, and for $b_0$ we get

$$b_0 = \int_{-\infty}^{\infty} h(t')p(t-t')dt' \bigg|_{t=n_0\Delta} = \int_0^{\Delta/2} \frac{1}{\tau} e^{-t'/\tau} \cdot 1 dt' = 1 - e^{-1/(2\tau')}. \qquad (20)$$

With Eq. (8) we then find $b_1 = 1 + a_1 - b_0 = \exp(-1/(2\tau')) - \exp(-1/\tau')$. The recurrence equation therefore is

$$y_n = e^{-1/\tau'} y_{n-1} + (1 - e^{-1/(2\tau')})x_n + (e^{-1/(2\tau')} - e^{-1/\tau'})x_{n-1}. \qquad (21)$$

This method also has a negligible implicit delay (not shown).

**2.8 Modified Tustin's method**

Below I will show that for implementing nonlinear feedback systems, a delay of $-\Delta/2$ is in fact favourable. One possibility is to use the ZOH for obtaining such a delay, but a modification of Tustin's method (the Trapezoidal rule discussed above) is at least as good, and has coeffients that are simpler to compute. Whereas the Trapezoidal rule has no appreciable implicit delay, because it weighs the present and previous inputs equally ($b_0 = b_1$), it can be given a $-\Delta/2$ delay by combining these weights to apply to the present input only:

$$y_n = [(\tau'-0.5)/(\tau'+0.5)]y_{n-1} + [1/(\tau'+0.5)]x_n. \qquad (22)$$

The method is evaluated along with the other methods in the remainder of this article, and will be shown to work very well for feedback systems. To my knowledge, this modification of Tustin's method has not been described in the literature before.

**3 Relationship between recursive schemes for first-order low-pass filters**

A Taylor expansion of the various forms of $-a_1$ gives

$$-a_1 = e^{-1/\tau'} = 1 - \frac{1}{\tau'} + \frac{1}{2\tau'^2} - \frac{1}{6\tau'^3} + ... \quad \text{for exponential Euler, ZOH, and FOH}, \qquad (23)$$

$$-a_1 = 1 - \frac{1}{\tau'} \qquad \qquad \qquad \text{for forward Euler}, \qquad (24)$$

$$-a_1 = \tau'/(\tau'+1) = 1/(1+1/\tau') = 1 - \frac{1}{\tau'} + \frac{1}{\tau'^2} - \frac{1}{\tau'^3} + ... \qquad \text{for backward Euler},\qquad(25)$$

$$-a_1 = (\tau'-0.5)/(\tau'+0.5) = (1-\frac{1}{2\tau'})/(1+\frac{1}{2\tau'}) = (1-\frac{1}{2\tau'})(1-\frac{1}{2\tau'}+\frac{1}{4\tau'^2}-\frac{1}{8\tau'^3}+...)$$
$$= 1 - \frac{1}{\tau'} + \frac{1}{2\tau'^2} - \frac{1}{4\tau'^3} + ... \qquad \text{for Trapezoidal and modified Tustin.}\qquad(26)$$

Compared to the theoretical exponential decline, Eq. (4), the exponential Euler, ZOH, and FOH are fully correct, the forward and backward Euler schemes are correct only up to the factor with $(1/\tau')$, whereas Trapezoidal and modified Tustin are correct up to the factor with $(1/\tau')^2$. The accuracy of the latter is related to the fact that $(\tau'-0.5)/(\tau'+0.5)$ is a first-order Padé approximation of $\exp(-1/\tau')$ (Bechhoefer 2005). Note that in the limit of $\tau >> \Delta$, all algorithms use approximately the same weight for the previous output sample, namely $1-1/\tau'$.

With respect to the weights acting on the input, the algorithms presented above can be divided into three groups, depending on the implicit delay they carry (see Table 1). If only the previous input sample is used (forward and exponential Euler), there is a delay of $\Delta/2$, if only the present input sample is used (backward Euler, ZOH, and modified Tustin's method) there is a delay of $-\Delta/2$, and if both the previous and present input samples are used (Trapezoidal and FOH), there is no delay. Below we will only analyze the groups with delays $-\Delta/2$ and 0.

The coeffients $b_0$ of the group with the phase advance (delay $-\Delta/2$) can be expanded as

$$b_0 = 1 - e^{-1/\tau'} = \frac{1}{\tau'} - \frac{1}{2\tau'^2} + \frac{1}{6\tau'^3} + ... \qquad \text{for ZOH},\qquad(27)$$

$$b_0 = 1/(\tau'+1) = \frac{1}{\tau'}\frac{1}{(1+1/\tau')} = \frac{1}{\tau'} - \frac{1}{\tau'^2} + \frac{1}{\tau'^3} - ... \quad \text{for backward Euler},\qquad(28)$$

$$b_0 = 1/(\tau'+0.5) = (\frac{1}{\tau'})/(1+\frac{1}{2\tau'}) = (\frac{1}{\tau'})(1-\frac{1}{2\tau'}+\frac{1}{4\tau'^2}-...)$$
$$= \frac{1}{\tau'} - \frac{1}{2\tau'^2} + \frac{1}{4\tau'^3} - ... \qquad \text{for modified Tustin,}\qquad(29)$$

where we find that ZOH and modified Tustin are more similar to each other than to backward Euler.

Finally, the coeffients of the FOH can be compared with those of Trapezoidal:

$$b_0 = 1 - \tau' + \tau'e^{-1/\tau'} = 1 - \tau' + \tau'(1-\frac{1}{\tau'}+\frac{1}{2\tau'^2}-\frac{1}{6\tau'^3}+...)$$
$$= \frac{1}{2\tau'} - \frac{1}{6\tau'^2} + ... \qquad \text{for FOH}\qquad(30)$$

$$b_0 = 0.5/(\tau'+0.5) = (\frac{1}{2\tau'})/(1+\frac{1}{2\tau'}) = (\frac{1}{2\tau'})(1-\frac{1}{2\tau'}+...)$$
$$= \frac{1}{2\tau'} - \frac{1}{4\tau'^2} + ... \qquad \text{for Trapezoidal}\qquad(31)$$

and

$$b_1 = \tau' - (1+\tau')e^{-1/\tau'} = \tau' - (1+\tau')(1 - \frac{1}{\tau'} + \frac{1}{2\tau'^2} - \frac{1}{6\tau'^3} + ...)$$
$$= \frac{1}{2\tau'} - \frac{1}{3\tau'^2} + ... \qquad \text{for FOH} \tag{32}$$

$$b_1 = 0.5/(\tau'+0.5) = \frac{1}{2\tau'} - \frac{1}{4\tau'^2} + ... \qquad \text{for Trapezoidal} \tag{33}$$

The coefficients start to differ in the factor with $(1/\tau')^2$. We will see in the examples below that FOH and Trapezoidal perform very similarly on concrete problems.

**4 Examples of nonlinear dynamic systems**

In this section I will provide several examples of nonlinear dynamic systems that are well suited to be simulated using autoregressive filters of the type discussed above. I will show for these examples how the systems can be rearranged to contain only static nonlinearities and first-order low-pass filters. Furthermore, I will compare the results of several of the algorithms presented above with an accurate numerical benchmark, and discuss the speed and accuracy of the various possibilities.

**4.1 Phototransduction: coupled nonlinear ODEs**

An example of a system where coupled nonlinear differential equations can be represented by a feedback system is the phototransduction system in the cones of the vertebrate retina. I will concentrate here on the main mechanism, which provides gain control and control of temporal bandwidth (van Hateren 2005). For the present purpose, a suitable form is given by

$$\dot{X} = 1/(1+C^4) - \beta X \tag{34}$$

$$\dot{C} = (X - C)/\tau_C. \tag{35}$$

The variable $\beta$ is linearly related to the light intensity, and can be considered as the input to the system. The variable $X$ represents the concentration of an internal transmitter of the cone, and can be considered as the output of the system because it regulates the current across the cone's membrane. The variable $C$ is an internal feedback variable, proportional to the intracellular $Ca^{2+}$ concentration.

We will now rewrite the equations such that they get the form of Eq. (2):

$$\tau_\beta \dot{X} = q/(1+C^4) - X$$
$$\text{with } \tau_\beta = 1/\beta \text{ and } q = 1/\beta \tag{36}$$

$$\tau_C \dot{C} = X - C. \tag{37}$$

By defining a time constant $\tau_\beta$ (actually not a constant, because it varies with $\beta$) and an auxiliary variable $q$, we see that both equations formally take a form similar to Eq. (2), where $q$ now has the role of input to Eq. (36), with the factor $1/(1+C^4)$ as a gain. We can thus represent these equations by the system diagram shown in Fig. 4A. The boxes containing a $\tau$ there represent unit-gain first-order low-pass filters. From the system diagram it is clear that the divisive feedback uses its own result after that has progressed through two low-pass filters and a static nonlinearity. The following describes the algorithm associated with Fig. 4A:

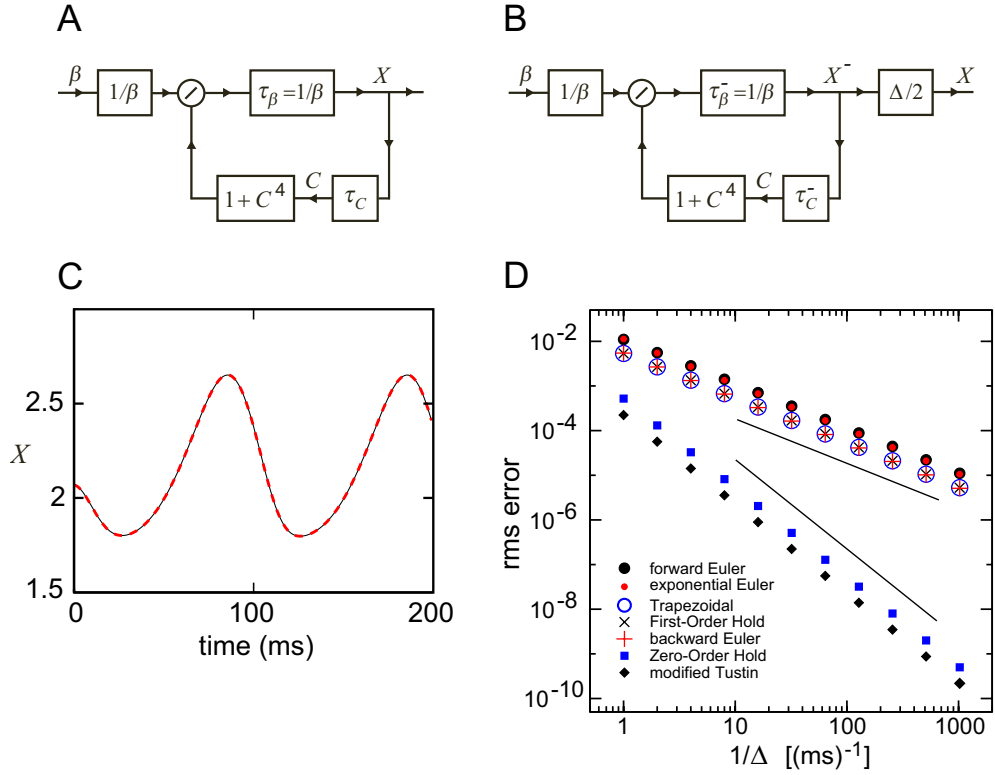

Figure 4. (A) System diagram of Eqs. (36) and (37). Boxes containing a 'τ' are unit-gain first-order low-pass filters, possibly depending on input or state variables (e.g., $\tau_\beta$ depends on $\beta$). The other boxes represent static nonlinearities given by the function definition inside the box. (B) Scheme equivalent to (A), where the required phase advance of one sample distance ($\Delta$) for the feedback is obtained by using two low-pass filters of type $\tau^-$ that each provide a $-\Delta/2$ delay (i.e., a $\Delta/2$ phase advance). The box to the right represents a $\Delta/2$ delay to compensate for the phase advance of $\tau_\beta^-$. (C) Thin black line: response $X$ of Eqs. (36) and (37), using $\tau_C$=3 ms, to $\beta = \beta_0(1+0.9\sin(2\pi f t))$ for $t \geq 0$ and $\beta = \beta_0$ for $t<0$, with $\beta_0$=0.025 (ms)$^{-1}$ and $f$=10 Hz, computed with Matlab ode45; dashed red line: result of filtering with the scheme of (B), with $\Delta$=1 ms and using the modified Tustin's method for $\tau^-$. (D) Root-mean-square (rms) error between the output when using the various recursive filters for the scheme of (B) and the result of ode45 at its maximum accuracy setting. Input as in (C). The thin straight lines are an aid for judging the scaling behaviour of the various methods, and have slopes of -1 and -2 in double-logarithmic coordinates.

- assume an initial steady state with $\beta = \beta_0$, and obtain initial values of all variables by solving (analytically or numerically) Eqs. (36) and (37) for $\dot{X}=0$ and $\dot{C}=0$
- repeat for each time step
    - read $\beta$ as input
    - compute $a_1$, $b_0$, and $b_1$ for $\tau_\beta = 1/\beta$, and update $X$ by low-pass filtering it, taking $(1/\beta)/(1+C^4)$ as input to the filter
    - use a precomputed $a_1$, $b_0$, and $b_1$ for $\tau_C$ to update $C$ by low-pass filtering it, taking $X$ as input to the filter
    - write $X$ as output

Note that $\tau_\beta$ is obtained from the current value of $\beta$. In principle, it might have been based partly on the previous value of $\beta$ as well, because $\beta$ changes in the interval between previous and current sample. However, for $\tau_\beta$ significantly larger than $\Delta$, this is expected to be a second-order effect, and the changing time constant is therefore treated in the simplest possible way, as described in the algorithm above.

Because at each time step only the result of *C* that was obtained at the previous time step can be used in the division by $(1+C^4)$, the feedback path would effectively get an (implicit) extra delay of $\Delta$ if calculated following this scheme. Such an extra delay will affect the results (and in extreme cases may lead to spurious oscillations), which can only be minimized by choosing $\Delta$ rather small. However, there is a way to alleviate this problem. As we have seen above, several of the autoregressive schemes have an implicit delay of $-\Delta/2$. Because there are two low-pass filters concatenated in the feedback loop, using such a scheme will produce a total delay of $-\Delta$, exactly compensating for the implicit delay $\Delta$ of the feedback. In other words, the divisor used at the point of divisive feedback will have the correct, current time. Because the forward low-pass filter, $\tau_\beta$, has a delay of $-\Delta/2$, we need to compensate that if we require that the output of the system has the right phase. (This may not always be necessary, especially not when the system is part of a larger system, where it would be more convenient to correct the sum of all delays at the final output.) The required delay of $\Delta/2$ can be approximated by linear interpolation, i.e., a recurrence equation $y_n = 0.5x_{n-1} + 0.5x_n$. The linear interpolation implies a slight low-pass filtering of the signal, and is therefore only accurate if the sampling rate is sufficiently high compared with the bandwidth of the signal. We can then replace the scheme of Fig. 4A by the one of Fig. 4B, where the symbol $\tau^-$ indicates that we are using filters with a $-\Delta/2$ delay (see Table 1).

How well do the recursive schemes of Section 2 perform on this problem? To evaluate that, the thin black line in Fig. 4C shows the response *X* of Eqs. (36) and (37) to a sinusoidal modulation of $\beta$, computed using the Matlab routine ode45 at high time resolution and high precision settings. The dashed red line shows the result when using the scheme of Fig. 4B with the modified Tustin's method used for $\tau^-$ with $\Delta=1$ ms. How the accuracy depends on $\Delta$ is evaluated in Fig. 4D, which shows the rms (root-mean-square) deviation from the ode45 benchmark as a function of $\Delta$, not only for the modified Tustin's method, but also for most of the other schemes. To get a fair comparison, the diagram of Fig. 4A was used for schemes with implicit delays 0 and $\Delta/2$, where for the latter an explicit delay of $-\Delta/2$ was added as a final stage. As is clear, the ZOH and especially the modified Tustin's method are superior. They scale more favourably as a function of $1/\Delta$, and for a given level of accuracy it is sufficient to use a $\Delta$ at least an order of magnitude larger than for the other schemes. They compute therefore at least an order of magnitude faster. Because of the simplicity and speed of computing the coefficients of the modified Tustin's method, this appears to be the scheme to be recommended for this type of feedback system. Note, however, that this scheme is only accurate when $\tau$ is at least a few times larger than $\Delta$ (Eqs. 26 and 29), and breaks down completely for $\tau'<1$ (with -$a_1$ even becoming negative for $\tau'<0.5$).

**4.2 Photopigment bleaching: dynamics on different time scales**

For an example of a stiff set of differential equations, we will look at the dynamics of photopigment bleaching in human cones (Mahroo and Lamb 2004, Lamb and Pugh 2004, van Hateren and Snippe 2007). For the present purpose, a suitable form of the equations is

$$\dot{R} = [I(1-B-R)-R]/\tau_R \qquad (38)$$

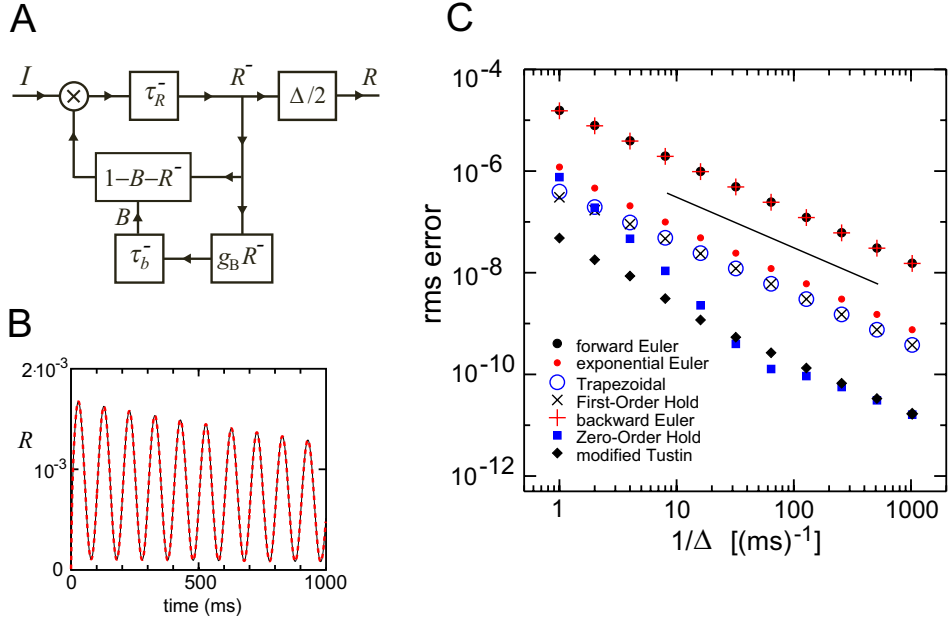

Figure 5. (A) System diagram of Eqs. (40) and (41). (B) Thin black line: response $R$ of Eqs. (40) and (41), using $\tau_R$=3.4 ms and $\tau_B$=25 s, to $I = 10^{-3}(1+0.9\sin(2\pi f t))$ for $t\geq 1$ ms, $I$=$10^{-5}$ for $t$<0, and $I$=$10^{-5}$+($10^{-3}$-$10^{-5}$)$t$ for $0\leq t<1$ ms, with $f$=10 Hz, computed with Matlab ode45; dashed red line: result of filtering with the scheme of (A), with $\Delta$=1 ms and using the modified Tustin's method for $\tau^-$. (C) Root-mean-square (rms) error between the various recursive filters used for the scheme of (A) and the result of ode45 at its maximum accuracy setting. Input as in (B). The thin straight line has a slope of -1 in double-logarithmic coordinates.

$$\dot{B} = R/\tau_R - \frac{0.2}{B+0.2}B/\tau_B. \tag{39}$$

Here $I$ is a (scaled) light intensity, $R$ is the (normalized) amount of photopigment excited by light, and $B$ the (normalized) amount of bleached photopigment. The rate by which excited pigment is bleached is governed by first-order kinetics ($1/\tau_R$), whereas the reconversion of bleached pigment to excitable pigment is governed by rate-limited dynamics (Mahroo and Lamb 2004): the second term in the right-hand-side of Eq. (39) is consistent with first-order kinetics for small $B$, but saturates for large $B$. Eqs. (38) and (39) form a stiff set of equations, because the time constants $\tau_R = 3.4 \cdot 10^{-3}$ s and $\tau_B = 25$ s differ substantially. Through the factor $(1-B-R)$, bleaching provides a slow gain control, controlling the sensitivity of the eye in bright light conditions.

Rewriting the equations into the form of Eq. (2) gives
$$\tau_R \dot{R} = I(1-B-R) - R \tag{40}$$

$$\tau_b \dot{B} = g_B R - B$$
$$\text{with} \quad \tau_b = \tau_B \frac{B+0.2}{0.2} \quad \text{and} \quad g_B = \tau_b / \tau_R. \tag{41}$$

This processing scheme is depicted in Fig. 5A, where $\tau_b$ and $g_B$ at time $t_n$ are derived from $B$ at time $t_{n-1}$. Note that the phase advance of $\tau^-$ is sufficient for the loop involving $\tau_b$, but only provides half of the required phase advance for the direct loop. Fig. 5B shows a benchmark calculation using ode45, and the result of using the scheme of Fig 5A with the modified Tustin's method. The stimulus $I$ steps at $t=0$ from $10^{-5}$ to a sinusoidal modulation around $10^{-3}$. Because an instantaneous step contains considerable power in its high-frequency components, using a recursive filter with a rather course $\Delta$ causes significant aliasing, which in this particular example would noticeably affect the response right after the step. To reduce the effect of aliasing, the step was assumed here to take 1 ms, that is, there is a linear taper between $t=0$ and 1 ms. Fig. 5C compares the rms error of the various schemes as a function of $\Delta$. Again, the ZOH and the modified Tustin's method perform best, despite the fact that there is no complete compensation of the feedback delay.

**4.3 Spiking neurons: Hodgkin-Huxley equations**

As a final example of a highly nonlinear system with fast dynamics, we will look at the Hodgkin-Huxley equations for spike generation (Hodgkin and Huxley 1952). Following the formulation by Gerstner and Kistler (2002, Chapter 2.2) these equations are given by Eqs. (42)-(45):

$$C\dot{u} = -g_{Na}m^3h(u-E_{Na}) - g_K n^4(u-E_K) - g_L(u-E_L) + I , \quad (42)$$

where $u$ is the membrane potential (in mV, defined relative to the resting potential), $C$ the membrane capacitance (taken as 1 μF/cm$^2$), the input variable $I$ is externally applied current, and the other terms represent membrane currents (consisting of a sodium, potassium, and leakage current). The membrane currents are given by the reversal potentials for the ions (in mV, defined relative to the resting potential: $E_{Na} = 115$, $E_K = -12$, and $E_L = 10.6$), by conductances (in mS/cm$^2$, $g_{Na} = 120$, $g_K = 36$, and $g_L = 0.3$), and by variables $n$, $m$, and $h$, describing the gating of the ion channels by the membrane potential

$$\dot{n} = \alpha_n(1-n) - \beta_n n \quad (43)$$

$$\dot{m} = \alpha_m(1-m) - \beta_m m \quad (44)$$

$$\dot{h} = \alpha_h(1-h) - \beta_h h . \quad (45)$$

The rate constants $\alpha$ and $\beta$ are functions of $u$, the form of which was determined empirically by Hodgkin and Huxley (1952): $\alpha_n = (0.1-0.01u)/[\exp(1-0.1u)-1]$, $\beta_n = 0.125\exp(-u/80)$, $\alpha_m = (2.5-0.1u)/[\exp(2.5-0.1u)-1]$, $\beta_m = 4\exp(-u/18)$, $\alpha_h = 0.07\exp(-u/20)$, and $\beta_h = 1/[\exp(3-0.1u)+1]$.

Rewriting the equations into the form of Eq. (2) gives

$$\tau_e \dot{u} = R_e(I + I_e) - u$$
$$\text{with} \quad I_e = g_{Na}m^3hE_{Na} + g_K n^4 E_K + g_L E_L \quad (46)$$
$$R_e = 1/(g_{Na}m^3h + g_K n^4 + g_L) \quad \text{and} \quad \tau_e = R_e C$$

$$\tau_n \dot{n} = n_\infty - n$$
$$\text{with} \quad \tau_n = 1/(\alpha_n + \beta_n) \quad \text{and} \quad n_\infty = \alpha_n/(\alpha_n + \beta_n) \quad (47)$$

$$\tau_m \dot{m} = m_\infty - m$$
$$\text{with} \quad \tau_m = 1/(\alpha_m + \beta_m) \quad \text{and} \quad m_\infty = \alpha_m/(\alpha_m + \beta_m) \quad (48)$$

$$\tau_h \dot{h} = h_\infty - h$$
$$\text{with} \quad \tau_h = 1/(\alpha_h + \beta_h) \quad \text{and} \quad h_\infty = \alpha_h/(\alpha_h + \beta_h) \quad (49)$$

This processing scheme is depicted in Fig. 6A. The feedback is partly additive (through the gated current $I_e$, which acts as a strong positive feedback during the rising phase of the spike, and as a negative feedback during the potassium-driven after-hyperpolarization), partly multiplicative (through the input resistance $R_e$, which drops considerably during the spike, and is the main cause of the absolute refractory period of the neuron), and partly through the time constant $\tau_e$, causing fast dynamics during the spike. Note that the system contains, for each of the three feedback variables, two low-pass filters in series ($\tau_e$ and the one belonging to either $n$, $m$, or $h$), thus we can fully utilize the phase advance of $\tau^-$ as in the example on phototransduction. Figures 6C and D show a benchmark calculation using ode45 of the response (black line) to a current input as shown in Fig. 6B. This stimulus is again tapered at the beginning to reduce aliasing. Some tapering is realistic, because normally the axon of a spiking neuron (where spiking starts) will not be driven by instantaneous current steps, but only by band-limited currents because of low-pass filtering by the cell body and dendrites. Figure 6C shows the result of using the scheme of Fig. 6A with Trapezoidal (obviously without the $\Delta/2$ processing block), and Fig. 6D with the modified Tustin's method. Fig. 6E compares the rms error of the various schemes as a function of $\Delta$. Again, the ZOH and the modified Tustin's method perform best. In particular the modified Tustin's method provides accurate results: even at a course $\Delta=1/2$ ms it misses no spikes in the example of Fig. 6, and the timing precision of the spikes is in the order of $0.1\Delta$. This contrasts with, for instance, a scheme like Trapezoidal, which needs $\Delta$ at least as small as 1/32 ms in order not to miss spikes, and has a timing precision of the spikes in the order of $10\Delta$.

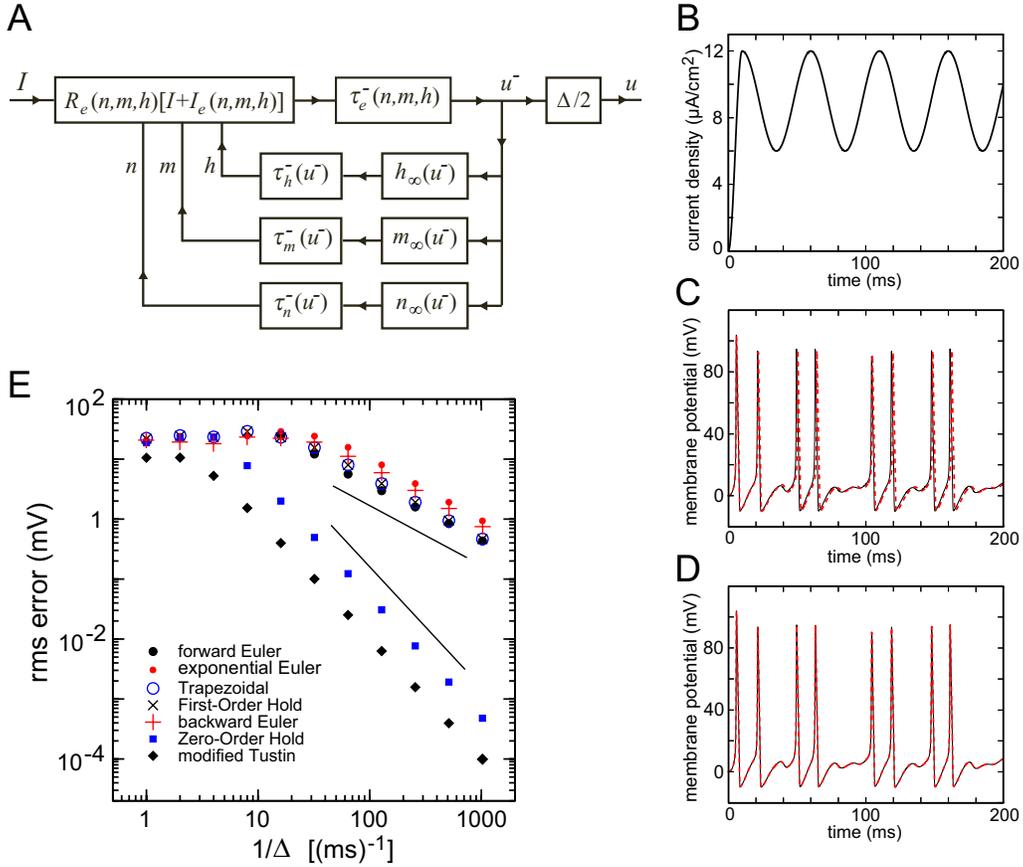

Figure 6. (A) System diagram of Eqs. (46) - (49). (B) Driving current density $I$, with $I=0$ for $t<0$, $I = I_0 \sin^2(0.5\pi t/t_0)$ for $0 \leq t < t_0$ ms, and $I = I_0(1 - 0.5\sin^2(0.5\pi f(t - t_0)))$ for $t \geq t_0$, with $t_0$=10 ms a taper, $f$=10 Hz, and $I_0$=12 μA/cm$^2$. (C) Thin black line: response $u$ of Eq. (46) to the stimulus defined at (B), computed with Matlab ode45; dashed red line: result of filtering with the scheme of (A), with $\Delta$=1/32 ms and using Trapezoidal for $\tau$. (D) Thin black line: as in (C); dashed red line: result of filtering with the scheme of (A), with $\Delta$=1/32 ms and using the modified Tustin's method for $\tau^-$. (E) Root-mean-square (rms) error between the various recursive filters used for the scheme of (A) and the result of ode45 at its maximum accuracy setting. Input as in (B). The thin straight lines have slopes of -1 and -2 in double-logarithmic coordinates.

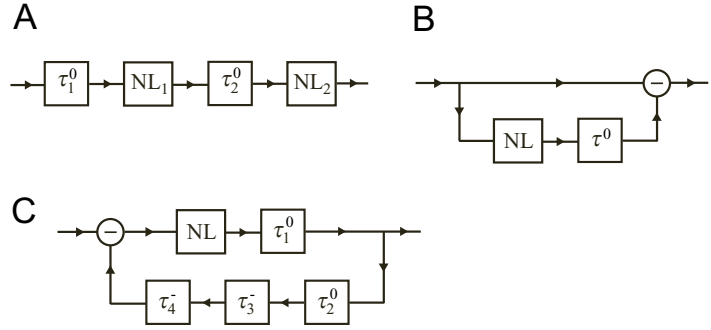

Figure 7. (A) Concatenation of low-pass filters and nonlinearities (NL), where zero-delay low-pass filters can be used. (B) In a feedforward loop as shown, a zero-delay low-pass filter should be used. (C) In a feedback loop, the total delay compensation needs to match the implicit delay $\Delta$ of the computational feedback scheme.

### 4.4 When to use $\tau^-$ or $\tau^0$

Two of the examples given above involve feedback with exactly two low-pass filters in the forward and backward branches of the feedback loop. For these schemes low-pass filters with phase advance are clearly useful. However, for other topologies this is not necessarily the case. Fig. 7 shows a few examples. When concatenating low-pass filters and static nonlinearities (Fig. 7A), zero-delay filters $\tau^0$ may be used, as an alternative to using $\tau^-$ and performing delay correction at a later stage. In a feedforward structure as shown in Fig. 7B, a zero-delay filter must be used. Similarly, if a feedback scheme contains more than two low-pass filters, some of the filters need to be zero-delay (Fig. 7C).

If a system contains a feedback loop with only one low-pass filter in either the feedforward or feedback branch, a filter $\tau^-$ can only provide half of the required phase advance. In those situations, as in the example on photopigment bleaching given above, it is still helpful to use $\tau^-$, in addition to making $\Delta$ sufficiently small. In principle, a phase advance (a delay of $-\Delta/2$) might be added by implementing it as a linear extrapolation $y_n = 1.5 x_n - 0.5 x_{n-1}$. However, I have not tested such a scheme, which might have stability problems.

Finally, if a feedback loop contains no low-pass filters at all, it is in fact identical to a static nonlinearity, and can usually be treated analytically, or via a precomputed look-up table.

### 4.5 Comparison with a 4th-order Runge-Kutta integration scheme

Although the present article focusses on simple autoregressive filters working on data with a given step size, it is interesting to compare the performance of the scheme with a standard integration method, such as 4th-order Runge-Kutta (RK4; Press et al. 1992). Figure 8 shows the results for RK4 and the modified Tustin's method, applied to a fairly complex model of the macaque retinal horizontal cell (van Hateren 2005). This model consists of cones connected to horizontal cells in a feedback circuit, and constitutes a cascade of a static nonlinearity, two nonlinear (divisive) feedback loops, and a subtractive feedback loop. All loops contain, in various configurations, low-pass filters and static nonlinearities. For details, such as parameter values and the differential equations involved, see van Hateren (2005).

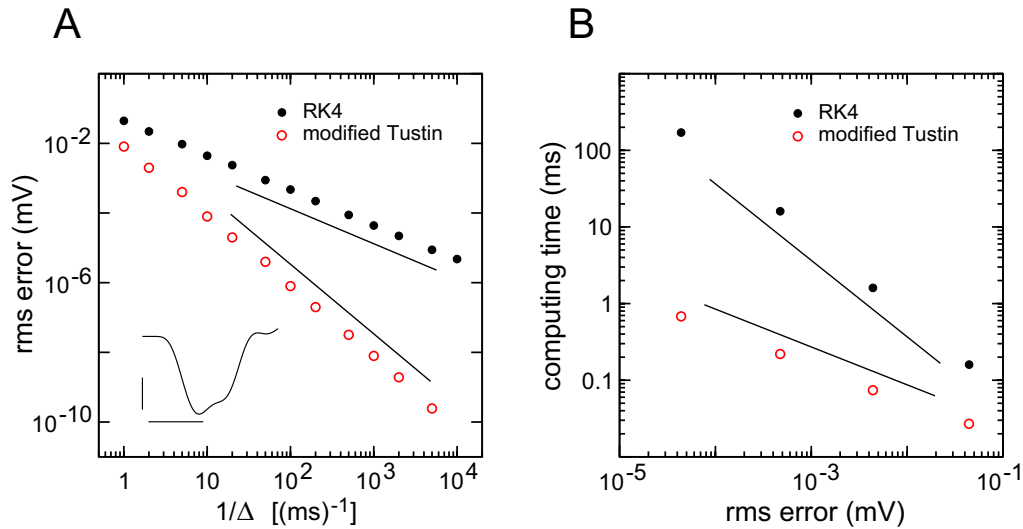

Figure 8. (A) Root-mean-square (rms) error of computing the response (inset, vertical bar = 2 mV) to a 40 ms light flash (horizontal bar inset) of the macaque retinal horizontal cell model of van Hateren (2005). Both a 4th-order Runge-Kutta scheme (RK4, fixed time step, routines rkdumb/rk4 of Numerical Recipes, Press et al. 1992; the input is an analytical block function according to the horizontal bar) and modified Tustin were implemented in a double-precision Fortran90 program (Intel compiler, Linux, 3.0 GHz Xeon). Errors are calculated relative to the result of modified Tustin at a time step $\Delta=0.1$ µs. The straight lines have slopes of -1 and -2 on double-logarithmic coordinates. (B) Computing times for RK4 and modified Tustin at matched rms error. For the four sets of data points the time steps $\Delta$ for (RK4, modified Tustin) are (1 µs, 70 µs), (10 µs, 230 µs), (0.1 ms, 0.7 ms), and (1 ms, 2.5 ms). Ratios of computing times are 250, 70, 20, and 6. The straight lines have slopes of -1 and -0.5 on double-logarithmic coordinates.

The inset in Fig. 8A shows the response of the model horizontal cell to a 40 ms light flash (horizontal bar) of contrast 2 given on a background of 100 td (see van Hateren 2005 for details on the stimulus). The vertical scale bar denotes 2 mV. This model was computed either using modified Tustin for the components (as in the examples in this article), or using RK4 for the entire set of differential equations. It should be stressed that this use of RK4 is different from the use of integrators, such as forward Euler, earlier in this article, where each low-pass filter was integrated separately. Here the RK4 algorithm is used, in the conventional way, on the entire model at once. All root-mean-square (rms) errors are calculated relative to the result of modified Tustin at a step size of 0.1 µs. Identical results were obtained when calculating all errors relative to RK4 at 0.1 µs, be it that errors then saturate at (i.e., do not go below) $4.7 \cdot 10^{-6}$ because of the limited accuracy of RK4 at 0.1 µs. Figure 8A shows the rms error of RK4 and modified Tustin. For all step sizes shown, modified Tustin outperforms RK4. The different scaling behaviour is indicated by the two lines with slopes of -1 and -2 on the double-logarithmic coordinates.

As argued by Morrison et al. (2007), in many situations the most interesting measure of performance of an integration method is the computing time required to achieve a given accuracy. This is shown in Fig. 8B for the two methods considered here. For this calculation the step size of modified Tustin was adjusted such that the accuracy of the result matched one of the RK4 calculations, and the corresponding computing times of the methods are plotted. Depending on accuracy, modified Tustin is typically 1-2 orders of magnitude faster than RK4. It should be noted that the calculation at the largest rms error already required a step size for modified Tustin (2.5 ms) that brought it well out of the range where the condition that the step size should be a few times smaller than $\tau$ (Eqs. 26 and 29) is valid, because the fastest low-pass filters in the model have time constants of 3-4 ms (van Hateren 2005). Nevertheless, even under these conditions modified Tustin is approximately 6 times faster than RK4 at the same accuracy.

**5 Discussion**

The fast recursive scheme presented in this article is particularly suited for situations where computing time is restrictive, for example when large arrays of neurons need to be computed. The scheme is fast, because each component is updated at each time step with only a few floating point operations. The examples given show that it is already quite accurate with fairly large time steps. It accomplishes this by computing feedback in a way that makes use of the fact that several autoregressive implementations of first-order low-pass filters produce an implicit phase advance of half a sample distance. The computational scheme is associated with a simple diagrammatic representation, that makes it relatively easy to get an intuitive understanding of the dynamics and of the processing flow, and allows for convenient symbolic manipulation (e.g., rearranging modules into equivalent schemes).

Because the $\tau$ of the low-pass filters may depend on input and system variables, the filter coefficients may require updating at each time step. This may constitute a significant part of the computational load. Fortunately, the coefficients for the Trapezoidal rule (for $\tau^0$) and the modified Tustin's method (for $\tau^-$) can be obtained with only a few floating-point operations. These schemes also give results at least as accurate as any of the other schemes, and therefore should be considered as first choice.

The present scheme is primarily intended for nonlinear filtering. It could be used for arbitrary linear filtering as well, because any linear filter can be approximated by a parallel arrangement of a number of low-pass filters with different weights and time constants. However, I have not tested how well the present scheme performs on such arrangements, and it seems likely that there are better ways to deal with arbitrary linear filters. One possibility is to use the matrix exponential (Rotter and Diesmann 1999), which is particularly suited when the signal consists of (or can be approximated by) point processes, as is common in calculating networks of spiking neurons. The matrix exponential can also be viewed as equivalent to a ZOH model and then needs a $\Delta/2$ compensation depending on whether it is used in a feedforward branch or is used as part of a nonlinear feedback branch. Another possibility is to use canned routines, like c2d in Matlab, that provide coefficients for a recursive discrete system corresponding to any rational continuous transfer function. For a linear filter that is part of a nonlinear feedforward loop, the c2d routines using FOH or Tustin's method are required, whereas ZOH is required when the linear filter is part of a feedback loop and a phase advance is wanted.

All calculations presented in this article were done with double precision arithmetic. For strongly stiff problems, such a precision is indeed necessary because of the large difference in time constants; the time step needs to be small enough to accommodate the shortest time constant, but such a short time step results in considerable error build-up in the processing of the largest time constant if single-precision arithmetic is used. However, I found that for the examples discussed in this article, single precision arithmetic already gives quite accurate results. This is of interest, because using single precision may accelerate computation, depending on processor architecture. Moreover, stream processors such as present-day GPUs may not yet support double-precision arithmetic (although double precision can be readily emulated, Göddeke et al. 2007, and GPUs with double precision are announced for the end of 2007).

I found that simulating the response of a large array of cones using the cone model of van Hateren and Snippe (2007), of which the examples of Sections 4.1 and 4.2 are part, provides performance one to two orders of magnitude higher on current GPUs than on current CPUs. Such performance is of interest for developing and testing models of the human retina (van Hateren, 2007) and also for using light adaptation in human cones as an algorithm for rendering and compression high-dynamic range video (van Hateren, 2006).

## Acknowledgments

I thank Sietse van Netten and Herman Snippe for comments on the manuscript.

**Table 1** Autoregressive filters approximating $\tau \dot{y} = x - y$ by $y_n = -a_1 y_{n-1} + b_0 x_n + b_1 x_{n-1}$, with sample distance $\Delta$, and $\tau' \equiv \tau/\Delta$

| Scheme | forward Euler | backward Euler | Trapezoidal rule | exponential Euler | Zero-Order Hold | First-Order Hold | modified Tustin's method |
|---|---|---|---|---|---|---|---|
| also known as | | | • Tustin's method <br> • Bilinear transformation <br> • Crank-Nicholson | • exponential integration | • step-invariant approximation <br> • Exact Integration | • ramp-invariant approximation <br> • triangular rule | |
| $-a_1$ (weight of $y_{n-1}$, previous output) | $(1-1/\tau')$ | $\tau'/(\tau'+1)$ | $(\tau'-0.5)/(\tau'+0.5)$ | $e^{-1/\tau'}$ | $e^{-1/\tau'}$ | $e^{-1/\tau'}$ | $(\tau'-0.5)/(\tau'+0.5)$ |
| $b_0$ (weight of $x_n$, present input) | - | $1/(\tau'+1)$ | $0.5/(\tau'+0.5)$ | - | $1-e^{-1/\tau'}$ | $1-\tau'+\tau'e^{-1/\tau'}$ | $1/(\tau'+0.5)$ |
| $b_1$ (weight of $x_{n-1}$, previous input) | $1/\tau'$ | - | $0.5/(\tau'+0.5)$ | $1-e^{-1/\tau'}$ | - | $\tau'-(1+\tau')e^{-1/\tau'}$ | - |
| implicit delay | $\Delta/2$ | $-\Delta/2$ | $0$ | $\Delta/2$ | $-\Delta/2$ | $0$ | $-\Delta/2$ |
| symbol | | | $\tau^0$ | | | | $\tau^-$ |
| remarks | can be unstable | | preferred choice for feedforward | | | | preferred choice for feedback |